\documentstyle[aps,prb,aps,graphicx,epsfig]{revtex}
\begin{document}


\draft{}
\title{Spin-dependent Transparency of Ferromagnet/Superconductor Interfaces}
\author{K. Xia,$^1$ 
P. J. Kelly,$^1$ 
G. E. W. Bauer,$^2$ and 
I. Turek$^3$}
\address{$^1$Faculty of Applied Physics and MESA$^{+}$ Research Institute,\\
University of Twente, P.O. Box 217, 7500 AE Enschede, The Netherlands,\\
$^2$Department of Applied Physics and DIMES, Delft University of Technology,\\
Lorentzweg 1, 2628 CJ Delft, The Netherlands,\\
$^3$Institute of Physics of Materials, Academy of Sciences of the Czech\\
Republic, CZ-616 62 Brno, Czech Republic}
\date{\today}
\maketitle

\begin{abstract}
Because the physical interpretation of the spin-polarization of a 
ferromagnet determined by point-contact Andreev reflection (PCAR) is
non-trivial, we have carried out parameter-free calculations of PCAR 
spectra based upon a scattering-theory formulation of Andreev 
reflection generalized to spin-polarized systems and a tight-binding
linear muffin tin orbital method for calculating the corresponding 
scattering matrices. PCAR is found to measure the spin-dependent
interface transparency rather than the bulk polarization of the 
ferromagnet which is strongly overestimated by free electron model 
fitting.
\end{abstract}


\vskip2pc]


The spin-polarization of a ferromagnet at the Fermi energy (not to be
confused with the total magnetic moment) is a parameter of crucial
importance entering all models of magneto-electronic phenomena such as
tunneling magnetoresistance (TMR). Although Moodera {\em et al.}'s TMR
experiments \cite{Moodera} could be interpreted in terms of bulk 
spin-polarizations using a very simple model due to Julliere 
\cite{Julliere}, subsequent experiments on a wider variety of 
materials cast doubt on the general validity of the model 
\cite{Sharma,deTeresa,LeClair}. They also led to the need to find
simpler ways of measuring spin-polarizations than the benchmark
determinations by Tedrow and Mersevey \cite{Tedrow-rep} using 
tunneling spectroscopy at ferromagnet$|$insulator$|$superconductor 
(S$|$I$|$F) junctions. These are difficult to make and characterize, 
especially for complex materials. Andreev reflection spectroscopy at
ferromagnet-superconductor (F$|$S) interfaces in ballistic point 
contacts (PCAR) has been proposed as a simpler and complementary method
to measure spin polarizations \cite{Buhrman,Soulen,Chien}. When the 
diameter of the point contact is smaller than the mean free path but
larger than the Fermi wavelength, transport is governed by the Sharvin
resistance \cite{Sharvin65}, appropriately modified to take into 
account the interface ``transparency''. In the Blonder-Tinkham-Klapwijk
(BTK) theory \cite{BTK}, this is done by representing the interface as
a planar $\delta $-function barrier between free-electron materials. 
The Bogoliubov-de Gennes equation \cite{Bogoliubov} for a 
normal-superconducting (N$|$S) interface is readily solved for this
model and the essential physics expressed in terms of a dimensionless
barrier strength $Z$ and the superconducting energy gap $\Delta $.
At the interface to a ferromagnetic material, the Andreev reflection
is suppressed by the difference between majority and minority 
electronic band structures \cite{deJong95}. Using the parabolic-band
Stoner-Wohlfarth model for ferromagnetic materials, BTK theory can
easily be generalized to spin-polarized systems. With only a single
additional parameter $P$, the spin polarization of the ferromagnet,
a good fit to experimental data is apparently achieved 
\cite{Buhrman,Chien}.

In spite of this, doubts have been raised about the definition and 
physical significance of the parameter $P$ derived from different 
experiments for all but 100\% polarized ferromagnets \cite{Mazin99}.
Apart from concerns about the relevant Fermi surface averaged quantity
to be used to describe different transport regimes, transport through
an interface should not depend solely on parameters characteristic of 
bulk materials. The representation of an interface between materials 
with quite different electronic structures as a $\delta $-function 
barrier can only be justified in terms of analytical simplicity. The
free-electron Stoner model does not describe itinerant ferromagnetism
in transition metals in an internally consistent manner and the 
probability of Andreev reflection is overestimated by such models; 
complex transition metal Fermi surfaces lead to a reduction of the 
phase space available to an electron to pair with a spin-flipped 
retro-reflected hole.

In this letter we present a generalized scattering formalism of Andreev
reflection \cite{Beenakker97,Lesovik97} to spin-polarized systems and
calculate the PCAR spectra with and without interface disorder. By 
basing our study on ab-initio calculations we can treat the full 
complexity of the transition metal band structure without introducing
any arbitrary parameters and make contact with experiments for a number
of systems of current experimental interest. We will be forced to 
conclude that the generalized BTK model omits an important aspect of the 
problem, basically because the bimodal distribution of the transmission
probabilities of real interfaces are inaccurately modelled using a 
$\delta $-function potential.

We start with the Bogoliubov-de Gennes \cite{Bogoliubov} equation for
the electron and hole wave functions of a superconductor 
\begin{equation}
\left( 
\begin{array}{cc}
H_{0\sigma }    & \Delta  \\ 
\Delta ^{\ast } & -H_{0-\sigma }^{\ast }
\end{array}
\right) \left( 
\begin{array}{c}
C_{e\sigma } \\ 
C_{h-\sigma }
\end{array}
\right) =\varepsilon \left( 
\begin{array}{c}
C_{e\sigma } \\ 
C_{h-\sigma }
\end{array}
\right) ,  \label{Hami}
\end{equation}
where $\sigma =\pm 1$. $H_{0\sigma}$ is the single-electron Hamiltonian
matrix for majority $(\sigma =1)$ and minority $(\sigma =-1)$ spins.
$C_{e\sigma (h-\sigma )}$ are the coefficient-vectors of the wave 
functions of electrons (holes) in some convenient basis. 
The excitation energy $\varepsilon $ is measured relative to the Fermi 
energy. Following Beenakker \cite{Beenakker97}, wave-function matching 
at the F$|$S interface is achieved by inserting between F and S 
a fictitious region in which S assumes its normal state (disregarding 
the proximity effect). At the F$|$N interface, the scattering of the 
Bloch states for electrons and holes can be written 
\begin{equation}
\left( 
\begin{array}{c}
C_{e\sigma }^{-}(F) \\ 
C_{e\sigma }^{+}(N) \\ 
C_{h-\sigma }^{+}(F) \\ 
C_{h-\sigma }^{-}(N)
\end{array}
\right)\!\! = \!\! \left( 
\begin{array}{cc}
\!\! S^{\sigma }(\varepsilon ) & \!\! 0                         \\ 
\!\! 0                   & \!\! S^{-\sigma \ast }(-\varepsilon )
\end{array}
\right) \!\! \left( 
\begin{array}{c}
C_{e\sigma }^{+}(F) \\ 
C_{e\sigma }^{-}(N) \\ 
C_{h-\sigma }^{-}(F) \\ 
C_{h-\sigma }^{+}(N)
\end{array}
\right) \label{SACTER1}
\end{equation}
where $+(-)$ denotes right(left) going waves and 
\begin{equation}
S^{\sigma }(\varepsilon ) =\left( 
\begin{array}{cc}
r_{11}^{\sigma }(\varepsilon ) & t_{12}^{\sigma }(\varepsilon ) \\ 
t_{21}^{\sigma }(\varepsilon ) & r_{22}^{\sigma }(\varepsilon )
\end{array}
\right)   \label{SCATTERING}
\end{equation}
is the normal-state scattering matrix. The subscript 1 refers to F, 
2 to N. At the (fictitious) N$|$S interface there is only Andreev 
scattering: 
\begin{eqnarray}
C_{e\sigma }^{-}(N) &=&\alpha C_{h-\sigma }^{-}(N) e^{i\phi} ,\\ 
C_{h-\sigma }^{+}(N)&=&\alpha ^{\ast}C_{e\sigma }^{+}(N) e^{-i\phi} , 
\label{Andr}
\end{eqnarray}
where for $|\varepsilon |<\Delta _{0}$,
$\alpha =\exp \left[ -i\arccos(\varepsilon /\Delta _{0})\right] $
describes the phase shift due to the penetration of the wave function 
into the superconductor and for $|\varepsilon |>\Delta _{0}$, 
$\alpha =(\varepsilon -{\rm sgn}\left( \varepsilon \right) 
\sqrt{\varepsilon ^{2}-\Delta _{0}^{2}})/\Delta _{0}$, 
$|\alpha |<1$. 
The incoming and reflected waves are related on the F side by 
\begin{equation}
\left( 
    \begin{array}{c}
    C_{e\sigma }^{-}(F) \\ 
    C_{h-\sigma }^{+}(F)
    \end{array}
\right) = \left( 
    \begin{array}{cc}
    R_{ee}^{\sigma } & R_{eh}^{\sigma } \\ 
    R_{he}^{\sigma } & R_{hh}^{\sigma }
    \end{array}
\right) \left( 
    \begin{array}{c}
    C_{e\sigma }^{+}(F) \\ 
    C_{h-\sigma }^{-}(F)
    \end{array}
\right). 
\label{RMATRIX}
\end{equation}
%
%
\begin{eqnarray}
R_{ee}^{\sigma }&=&r_{11}^{\sigma }(\varepsilon ) \nonumber \\
                &+&\alpha^2 t_{12}^{\sigma}(\varepsilon)
                r_{22}^{-\sigma \ast }(-\varepsilon )
\frac{1}{1-\alpha^{2}r_{22}^{\sigma }(\varepsilon )
                     r_{22}^{-\sigma \ast }(-\varepsilon )}
                  t_{21}^{\sigma }(\varepsilon )  
\label{REE}
\end{eqnarray}
\begin{equation}
R_{he}^{\sigma }=\alpha ^{\ast }e^{-i\phi }
       t_{12}^{-\sigma \ast}(-\varepsilon )
    \frac{1}{1-\alpha ^{2}r_{22}^{\sigma }(\varepsilon)
                          r_{22}^{-\sigma \ast }(-\varepsilon )}
       t_{21}^{\sigma }(\varepsilon )
\label{REH}
\end{equation}
are the reflection coefficients for the total system. In terms of the
spectral conductance 
\begin{equation}
G_{FS}(\varepsilon )=\frac{e^{2}}{h}
\sum_{\sigma =\pm 1}
Tr(1 - R_{ee}^{\sigma }R_{ee}^{\sigma \dagger } +
       R_{he}^{\sigma }R_{he}^{\sigma \dagger }),  \label{GFS}
\end{equation}
and the Fermi-Dirac distribution function $f$, the current-voltage 
relation is 
$eI(V)=\int d\varepsilon \lbrack f(\varepsilon )-f(\varepsilon
+eV)]G_{FS}(\varepsilon )$. 
The charge transport in the presence of Andreev reflection is thus
expressed in terms of the scattering matrix for the normal state. This
expression is the spin polarized generalization of Beenakker's formula 
\cite{Beenakker97} and reduces to the BTK equivalent if we use the
scattering matrix of a $\delta $-function potential and the parabolic 
band model. Note that the expressions also hold for energies above
the superconducting gap.

The parameter-free calculation of the reflection matrices \cite{Xia01} 
and the conductance (\ref{GFS}) is based on the surface Green's 
function method \cite{Turek} implemented with a tight-binding linear
muffin tin orbital basis\cite{Andersen85}. Because a minimal basis set
is used, we are able to carry out calculations for large lateral supercells 
and model disorder very flexibly within such supercells without using any adjustable parameters.
The electronic structure is determined self-consistently within the 
local spin density approximation. For disordered layers the potentials
are calculated using the layer CPA approximation 
\cite{Turek}. 

We focus our attention here on the Cu/Pb, Co/Pb and Ni/Pb systems
investigated experimentally in \cite{Buhrman} which were grown with
an {\it fcc}(111) orientation in point contacts. We encounter the
practical difficulty that the lattice constants of Pb are quite 
different from those of Cu, Co and Ni. Lattice matching is modelled
using lateral supercells containing $4 \times 4$ Cu (Co or Ni) 
atoms on one side of the interface and $3 \times 3$ Pb atoms/monolayer
on the other side or $7 \times 7$ Cu, Co or Ni and $5 \times 5$ Pb 
atoms/monolayer which gives better matching but is computationally 
more expensive. Disorder is taken into account by introducing a two 
monolayer-thick 50\% interface alloy \cite{Xia01}. The results are not
very sensitive to moderate variations in the alloy concentration in 
this range. Experiments \cite{Buhrman} have been carried out at 
slightly elevated temperatures to suppress a possible proximity effect.
Our computations are therefore carried out using the corresponding
values of kT in the Fermi function: 4.2~K for Cu/Pb, 1.4~K for 
Co/Pb and 2.5~K for Ni/Pb. The only parameter in the calculations is 
the superconducting energy gap $\Delta_{\rm Pb}$ which is taken
from experiment and is very close to the bulk value.

\begin{figure}[b]
\includegraphics*[scale=0.6,angle=-90]{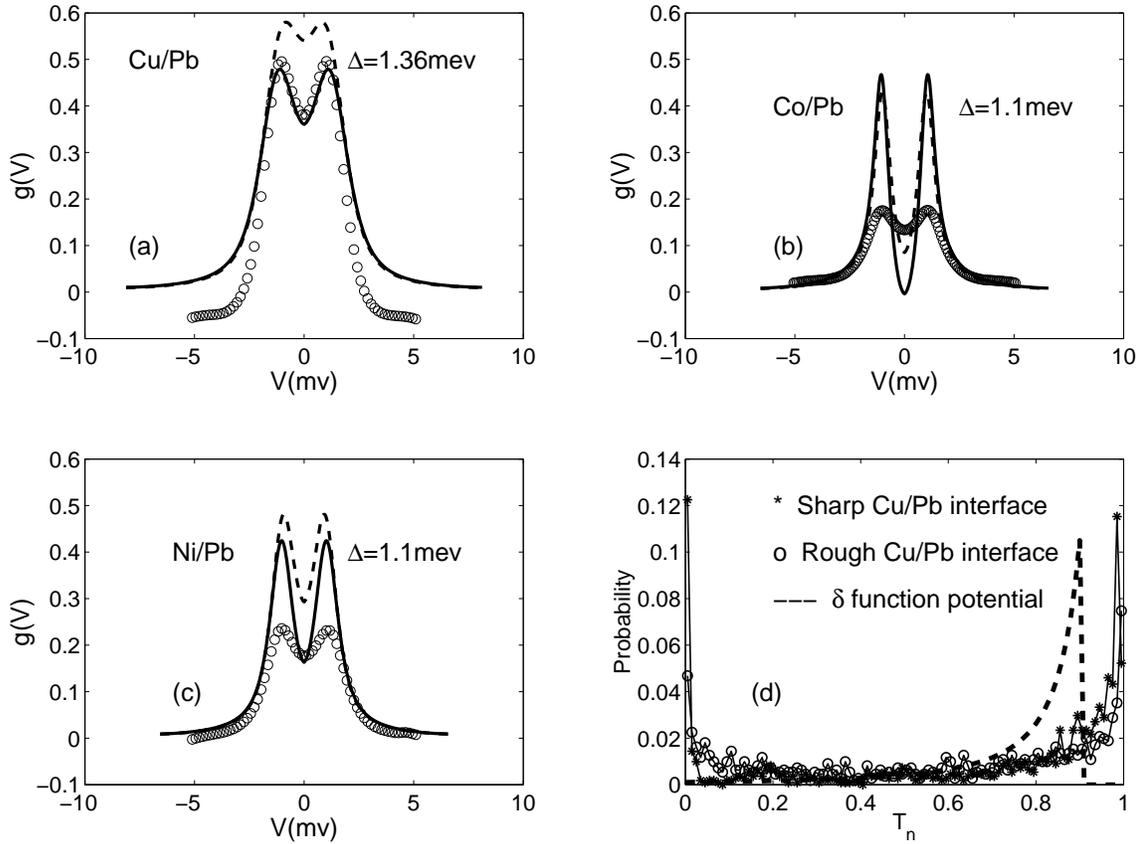}
\caption{(a)-(c)Calculated $g(V)$ curves for disordered (solid line) 
and clean interfaces (dashed line) compared with the Cornell group's experimental results (o).
(a) Cu-Pb, (b) Co-Pb, (c) Ni-Pb.
(d) Distribution function of the transmission matrix eigenvalues 
of a clean (x) and disordered (o) Cu/Pb interface. The distribution 
function of the BTK model with $Z=0.3$ is plotted as a dashed line
for comparision. }
\label{Fig1}
\end{figure}

For an ideal N$|$S interface Andreev reflection leads to a doubling
of the conductance. In Table~\ref{Tab1} the zero bias values 
$G_{FS}(0)/G_{FN}(0)$, calculated for clean and alloy interfaces, are
compared with the experimental values of Ref.\cite{Buhrman}. Suppression
of the Andreev reflection is apparent in all three cases but is 
smallest for non-ferromagnetic Cu. Even there, however, bandstructure
mismatch between Cu and Pb gives rise to a significant reduction of the
interface transparency and suppression of AR. The agreement with 
experiment is remarkably good and only in the case of Co/Pb is there
any indication of a discrepancy. This picture changes when we compare
the calculated and experimental spectra in Fig. \ref{Fig1} shown in 
terms of $g(V)\equiv \left[ G_{FS}(V)-G_{FN}(0)\right] /G_{FN}(0)$ 
where $G_{FN}(0)$ is the normal state conductance and $G_{FS}(V)$ is 
the differential conductance for an F$|$S interface at finite 
temperature and finite bias.

Let us first focus on the paramagnetic Cu/Pb point contact, 
Fig.~\ref{Fig1}(a). Whereas results for a specular interface deviate 
significantly from the measured data, the decreased conductance of a
rough interface leads to essentially perfect agreement with experiment.
Aspects of the problem not taken into account, like spin-orbit 
scattering and strong-coupling superconductivity in Pb are apparently
not very important.

The agreement with the PCAR spectra for the ferromagnetic systems 
(Figs. \ref{Fig1}(b) and (c)) is not satisfactory close to the band
edge of quasi-particle excitations, where the theoretical results 
appear to strongly exaggerate the Andreev reflection \cite{Taddei01}.
Since we have no parameters with which we could improve the agreement
the deviations should be sought in deficiencies of the model. 
For example, the close proximity of the ferromagnet might smear out 
the peaked density of states of the superconductor at the band edges 
\cite{Halterman}. Whereas this point deserves attention, we focus in
the following on the zero-bias conductance.

The spin-polarization of the interface conductance,
$P=(G_{maj}-G_{min})/(G_{maj}+G_{min})$, will in general differ
from that of the bulk ferromagnetic material. For example,
for Ni/Pb, $P=0.5\%$ for a specular interface and $-1.4\%$ 
for a rough interface. This should be compared to the BTK fit obtained
in Ref. \cite{Buhrman} of $P_{BTK}=32\%$. Our prediction of a small
spin-dependence of the interface conductance could be tested directly
by measuring the magnetoresistance of a Ni/Pb/Ni trilayer (or 
multilayer) in the current-perpendicular-to-the-plane configuration
with the layer thickness chosen sufficiently thin that the total
resistance is dominated by the interface contribution. Using the
expression given by Schep {\it et al.}\cite{schep1} for diffuse 
systems, we predict interface resistances of 
$R_{maj}=0.97\ f\Omega m^{2}$ and 
$R_{min}=1.27\ f\Omega m^{2}$ for a rough Ni/Pb (111) interface.

We emphasize the difference between the single-channel BTK model 
and the multichannel scattering theory generalization by focusing
on the zero-voltage, zero-temperature conductance (\ref{GFS}) for a 
non-magnetic system: 
\begin{equation}
\! G_{NS} \! = \! \frac{4e^2}{h} \sum_n \frac{{T_n}^2}{(2-T_n)^2}
=\frac{4e^2}{h} \! \int \! dT\rho (T)\frac{T^2}{(2-T)^2}
\end{equation}
is expressed in terms of the eigenvalues of the transmission matrix,
$T_n$, and the sum is over the propagating channels \cite{Beenakker97}.
For the Cu/Pb interface, the distribution function of transmission
matrix eigenvalues, $\rho (T)$, is shown in Fig.~\ref{Fig1}(d). It turns
out to be a bi-modal function with a clear dependence on 
interface roughness and on the constituent materials. Obviously, 
$\rho (T)$ strongly affects the Andreev reflection probability with 
the main contribution coming from the highly transmitting part, whereas
the average conductance is much more affected by the low transmission
channels. Using the Andreev conductance to derive an average 
transmission is therefore misleading. For a $\delta $-function 
potential and free electrons the transmission probability as a function
of the in-plane wave vector has the form 
$T_{k_{\parallel }}= 1/
     \left(    1 + \frac{Z^2}{1-(k_{\parallel}/k_F)^2}    \right)$.
The distribution function for $Z=0.3$, the value which must be taken
to reproduce the average conductance, is plotted as a dashed curve in 
Fig.~\ref{Fig1}(d) and is seen to be qualitatively incorrect compared to
the realistic distributions. 

For the F$|$S interface the situation is more complicated. Since Cooper
pairs are phase sensitive, the conductance is governed by the full 
scattering matrix, {\em \i.e.} by the scattering amplitudes and not 
just the transmission probabilities. We therefore cannot use a single 
distribution function to characterize the interface because the Andreev 
reflection probability contains terms like 
$1 - \alpha ^{2} r_{22}^{\sigma }(\varepsilon )
                 r_{22}^{-\sigma \ast}(-\varepsilon )$ 
which cannot be expressed by $T_{\uparrow }$ and $T_{\downarrow }$. 
Still, it should be clear that the deficiencies of the $\delta $ 
function barrier potential will not vanish.

In summary we calculate the PCAR spectra of Co/Pb, Ni/Pb, and Cu/Pb 
point contacts and the spin-polarization of Co/Pb, Ni/Pb interfaces 
from first principles which allows us to remove what is a dubious
approximation made when applying the BTK model to transition metals.
We find good agreement with experimental results by the Cornell group
for the full spectrum of Cu/Pb and for the zero 
bias conductance in the ferromagnetic systems. Qualitative differences
are revealed by comparing our results with that of a simple BTK model.
The physical significance of parameters obtained by fitting experiments
with this simple model is doubtful.

This work is supported by the Stichting FOM, 
the NEDO International Joint Research Grant Program
``Nano-magnetoelectronics'', 
the Grant Agency of the Czech Republic (202/00/0122), 
the European Commission's TMR Research Network on ``Interface Magnetism''
(contract No. FMRX-CT96-0089) and 
RT Network ``Computational Magnetoelectronics'' 
(contract No. HPRN-CT-2000-00143).
We acknowledge fruitful discussions with W. Belzig, A. Brataas, 
Yu. Nazarov, and J. Kudrnovsky and thank R.A. Buhrman for providing
us with the experimental data shown in Figure 1.

\begin{table}[tbp]
\begin{center}
\begin{tabular}{cccc}
$G_{FS}(0)/G_{FN}(0)$ & Cu/Pb & Ni/Pb & Co/Pb \\ \hline
Clean Interface       & 1.54  & 1.29  & 1.08  \\ 
Alloy Interface       & 1.36  & 1.16  & 1.00  \\ 
Experiment            & 1.38  & 1.18  & 1.13  \\ 
\end{tabular}
\end{center}
\caption{Zero bias suppression of Andreev reflection compared with the
experimental values obtained from Ref.~\protect\onlinecite{Buhrman}.}
\label{Tab1}
\end{table}

\draft{}

\end{document}